\documentclass[a4paper,11pt]{article}
\pdfoutput=1 

\usepackage{jinstpub} 

\usepackage{subcaption}

\usepackage{lineno}

\title{\boldmath FPGA-based real-time data processing for accelerating reconstruction at LHCb}

\author[a,b,1]{F. Lazzari,\note{Corresponding author.}}
\author[c,d]{W. Baldini,}
\author[a,e]{G. Bassi,}
\author[f]{A. Contu,}
\author[g]{M. Dorigo,}
\author[a,h]{R. Fantechi,}
\author[i,j]{L. Giambastiani,}
\author[a,e]{M. J. Morello,}
\author[a,h]{G. Punzi,}
\author[a,h]{M. Sticchi,}
\author[k]{G. Tuci}

\affiliation[a]{INFN Sezione di Pisa,\\Largo B. Pontecorvo 3, Pisa, Italy}
\affiliation[b]{Universit\`a degli Studi di Siena,\\Via Banchi di Sotto 55, Siena, Italy}
\affiliation[c]{INFN Sezione di Ferrara,\\Via Saragat 1, Ferrara, Italy}
\affiliation[d]{Universit\`a degli Studi di Ferrara,\\Via Ludovico Ariosto 35, Ferrara, Italy}
\affiliation[e]{Scuola Normale Superiore,\\Piazza dei Cavalieri 7, Siena, Italy}
\affiliation[f]{INFN Sezione di Cagliari,\\S.P. per Sestu, Monserrato, Italy}
\affiliation[g]{INFN Sezione di Trieste,\\Via Valerio 2, Trieste, Italy}
\affiliation[h]{Universit\`a di Pisa,\\Lungarno Pacinotti 43, Pisa, Italy}
\affiliation[i]{INFN Sezione di Padova,\\Via Marzolo 8, Padova, Italy}
\affiliation[j]{Universit\`a di Padova,\\Via 8 Febbraio 2, Padova, Italy}
\affiliation[k]{University of Chinese Academy of Sciences,\\No.19(A) Yuquan Road, Shijingshan District, Beijing, P.R.China}

\emailAdd{federico.lazzari@cern.ch}

\abstract{In Run-3, beginning in 2022, the LHCb software trigger will start reconstructing events at the LHC average crossing rate of 30 MHz. Within the upgraded DAQ system, LHCb established a testbed for new heterogeneous computing solutions for real-time event reconstruction, in view of future runs at even higher luminosities.
One such solution is a highly-parallelized custom tracking processor (“Artificial Retina”), implemented in state of the art FPGA devices connected by fast serial links.
We describe the status of a realistic prototype for the reconstruction of pixel tracking detectors that will run on real data during Run-3.}


\collaboration[c]{on behalf of LHCb collaboration}

\proceeding{Topical Workshop on Electronics for Particle Physics - TWEPP2021\\
  20 - 24 September, 2021}

\begin{document}
\maketitle
\flushbottom

\section{Introduction}
\label{sec:intro}
With the the slowing down of Moore's law, HEP experiments are looking at heterogeneous computing solutions as a way to manage ever-increasing data flows and complexity.
LHCb is at the frontier of these developments due to its specific physics needs, calling for the full software reconstruction of events in real-time at the LHC average rate of 30 MHz.
LHCb has adopted a GPU-based solution for the first stage of the High Level Trigger (HLT1) for Run-3. Further computing enhancement are being sought for Upgrade-II, where the luminosity will increase by a factor from 5 to 10 respect to Run-3.
To this purpose, a coprocessor testbed has been established, to allow parasitical testing of new processing solutions in realistic DAQ conditions during the 2022 run.

One such solution under development is a highly-parallelized custom tracking processor based on the ``Artificial Retina''.
The ``Artificial Retina'' architecture~\cite{retina} takes advantage of FPGA parallel computational capabilities, by distributing the processing of each event over an array of FPGA cards, interconnected by a high-bandwidth ($\sim$15 Tb/s) optical network. This is expected to allow operation in real-time at the full LHC collision rate, with no need for time-multiplexing or extra buffering due to its low latency ($< 1$ $\mu$s).

\begin{figure}[htbp]
  \centering
  \includegraphics[width=0.4\linewidth]{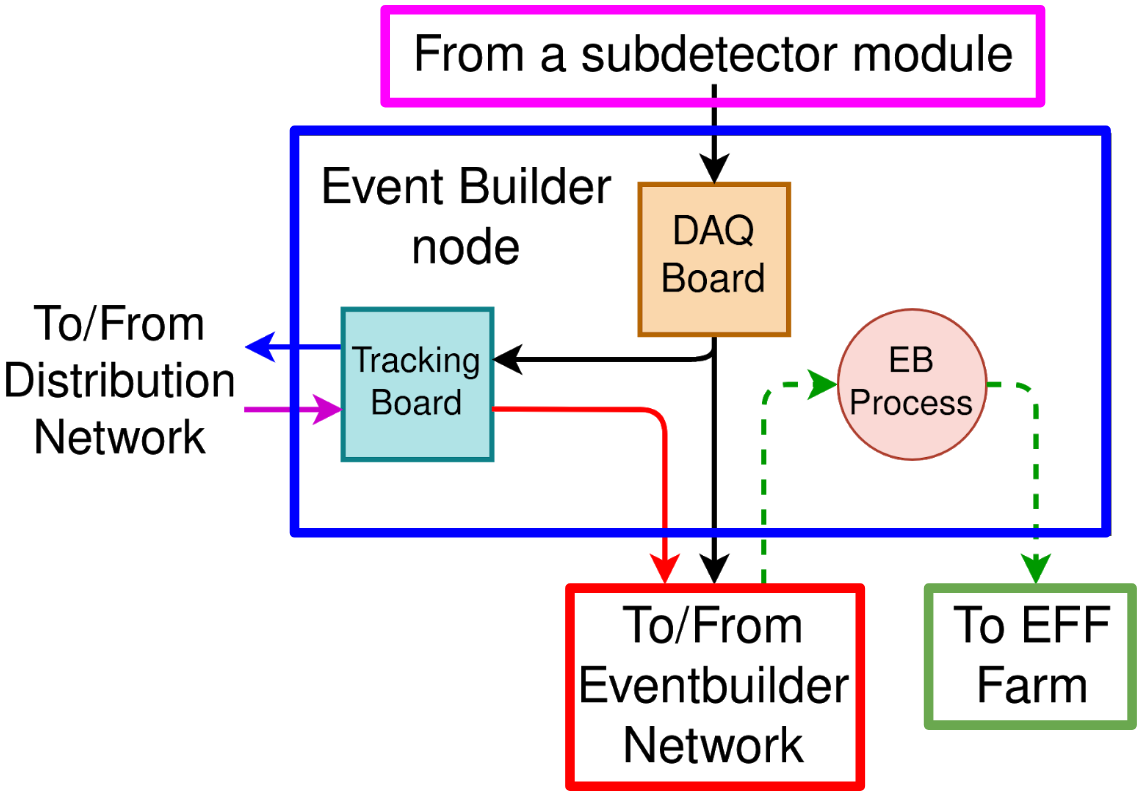}
  \caption{Integration of the tracking system in the LHCb Event Builder.}
  \label{fig:integration}
\end{figure}

Achieving this level of performance will allow it to be integrated into the DAQ chain of the experiment.
The Tracking Boards, hosting the ``Artificial Retina'' FPGA, might be installed in the Event Builder nodes paired to the DAQ Boards of desired sub-detector (Figure~\ref{fig:integration}), and provide tracks to the Event Builder like a virtual sub-detector.
These data can be used as seeds by the High Level Trigger to find tracks and perform trigger decisions with less computational effort than needed by starting from the raw detector data.

We are building the first realistic prototype of this system.
The prototype is designed to process data from a portion of the VELO pixel detector.
The VELO detects charged particles in the region closest to the interaction point, aiming at reconstructing primary and secondary vertexes with a spatial resolution smaller than typical decay lengths of $b$- and $c$-hadrons in LHCb ($c\tau \sim$ 0.01 -- 1~cm), in order to discriminate between them.
It consists of 52 modules positioned along the beam axis, 38 of which are located downstream of the nominal interaction point. Each module is read-out by a DAQ Board.
That is compact enough to be processed by a small system, and yet accounts for a significant fraction ($\sim$25 \%) of HLT1 computational load.

The implementation of this technology for other LHCb track detectors could enhance the physics potential of LHCb as early as at the start of Run-4 of the LHC, by expanding its HLT1 trigger capability with the inclusion of long-lived tracks like the $K^0_S$ and the $\Lambda^0$~\cite{down}.


\section{The ``Artificial Retina'' architecture}
\label{sec:artificial_retina}

\begin{figure}[htbp]
  \centering
  \begin{subfigure}[b]{0.3\textwidth}
    \centering
    \includegraphics[width=\textwidth]{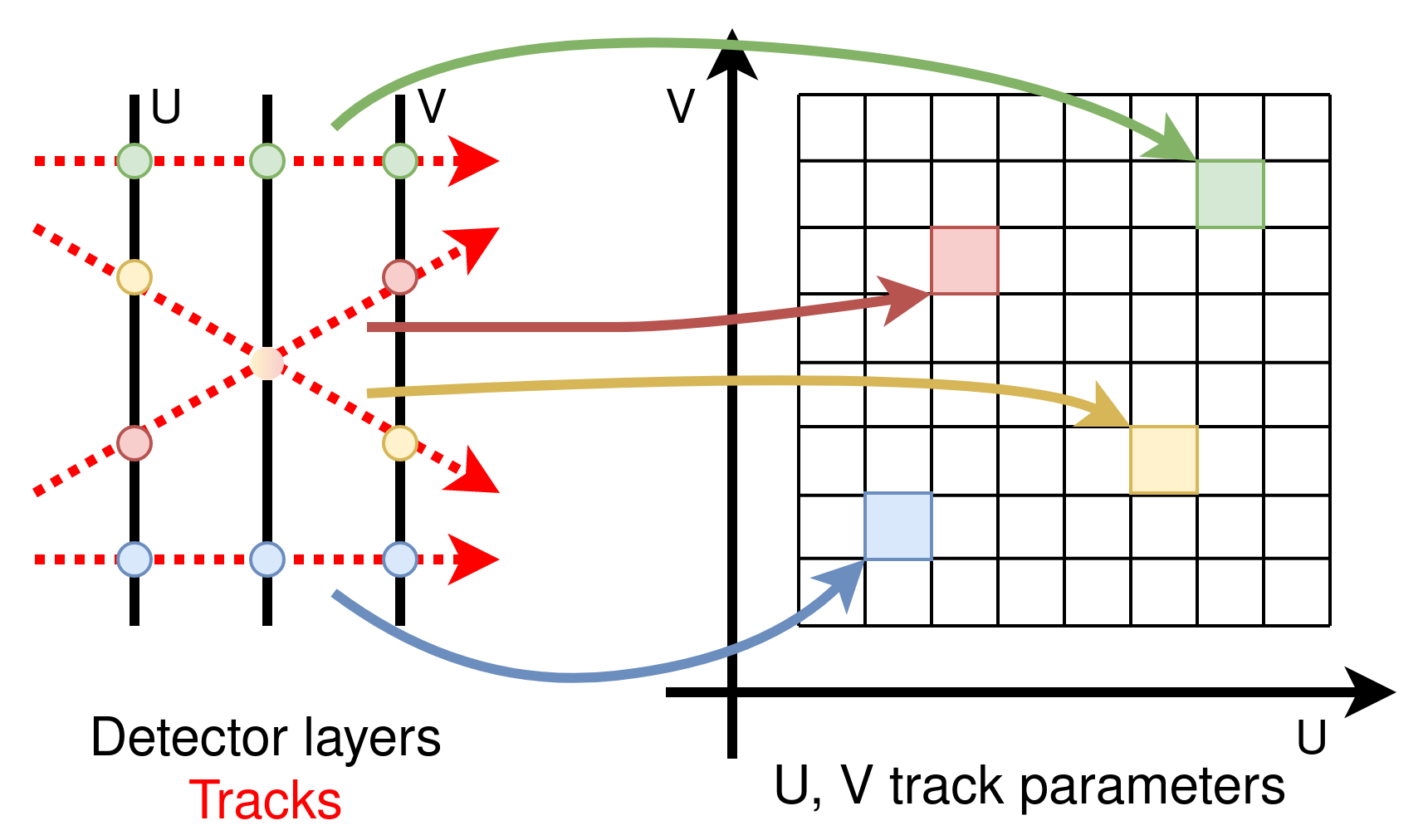}
    \caption{}
    \label{fig:retina_step_mapping}
  \end{subfigure}
  \quad
  \begin{subfigure}[b]{0.3\textwidth}
    \centering
    \includegraphics[width=\textwidth]{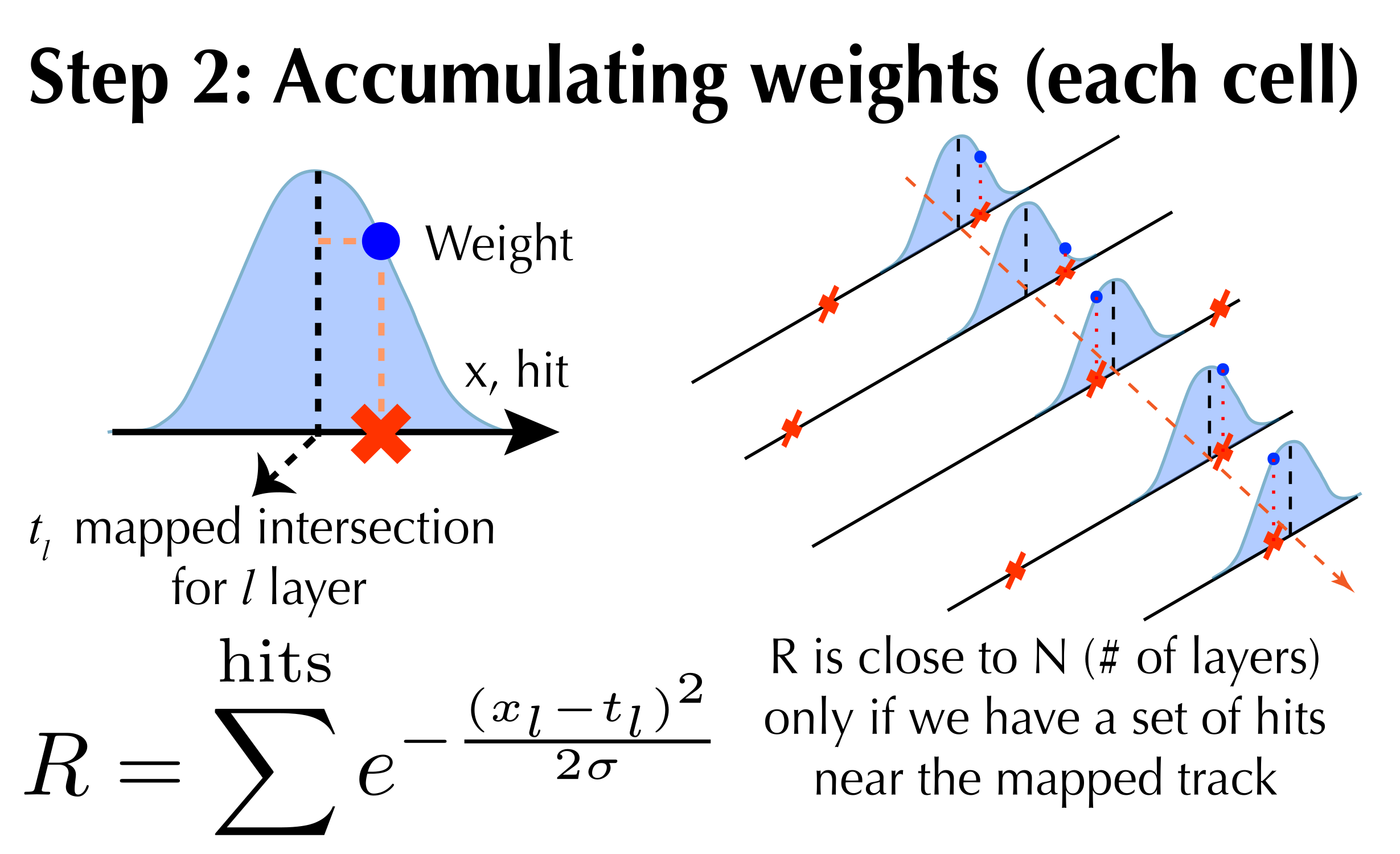}
    \caption{}
    \label{fig:retina_step_2}
  \end{subfigure}
  \quad
  \begin{subfigure}[b]{0.3\textwidth}
    \centering
    \includegraphics[width=\textwidth]{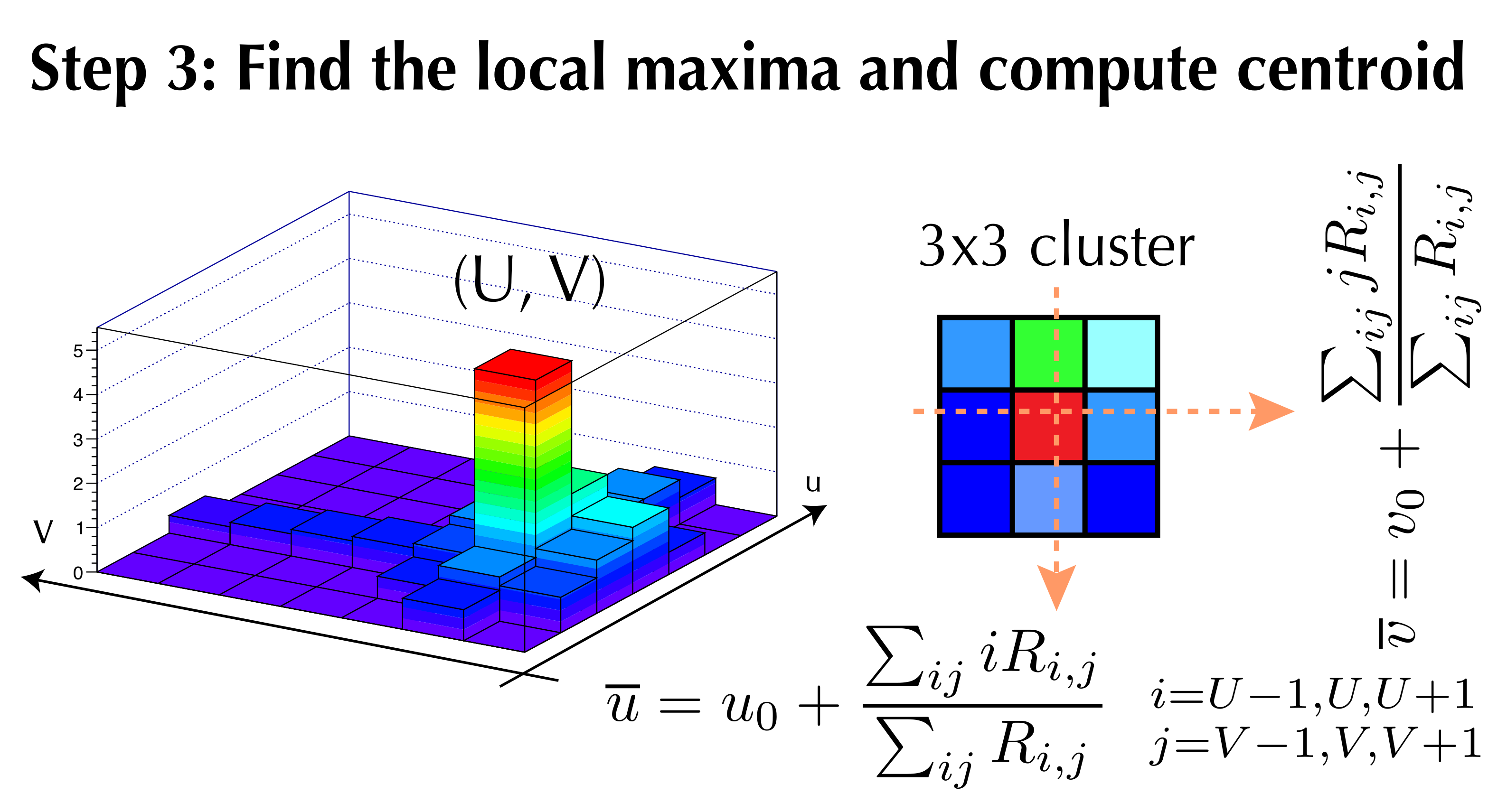}
    \caption{}
    \label{fig:retina_step_3}
  \end{subfigure}
  \caption{Track reconstruction steps with the ``Artificial Retina'' architecture.}
  \label{fig:retina_step}
\end{figure}

In the ``Artificial Retina'' architecture, the track parameters space is represented by a matrix of cells (Figure~\ref{fig:retina_step_mapping}).
The center of each cell corresponds to a reference track in the detector that intersects the layers in specific spatial points called receptors.

Each cell computes a weighted sum of hits nearby the reference track. The weights are proportional to the distance between the hits and the receptor. The resulting sum approaches the number of subdetector layers as the set of hits gets closer to the reference track (Figure~\ref{fig:retina_step_2}).

Subsequently, cells search local maxima in the matrix of weighted sums. Track parameters are reconstructed by interpolating the responses of cells nearby the local maxima (Figure~\ref{fig:retina_step_3}).

To overcome FPGA size limitations without increasing latency, cells are spread over several chips that work in parallel on the same event.
Usually hits provided by a DAQ Board belong to a limited region of the sub-detector. In order to perform tracking we need to exchange data between FPGAs.
Modern FPGAs have numerous high-bandwidth transceivers (XCVRs), that can be used to implement optical serial links between boards.
A custom network redistributes hits by track parameter coordinates, performing something similar to a ``change of reference system''.
Furthermore, by using Lookup Tables (LUTs) the Distribution Network delivers to each cell only the hits close to its receptors, resulting in a smaller number of hits to be processed by the cell. This allows the system to reach higher throughput.

We designed a modular Distribution Network, with the dispatcher as the basic block.
It has two inputs and two outputs, being able to send any input to any output (even both) according to the LUT routing scheme. We call an entity composed of several dispatchers a switch.
Combining a sufficient number of dispatchers it is possible to build a Distribution Network with the desired number of inputs and outputs.
The network is implementable within the same array of FPGAs performing the tracking, in separate and independent locations of the chip.

\section{Velo Clustering}
\label{sec:clustering}
\begin{figure}[htbp]
  \centering
  \includegraphics[width=0.8\linewidth]{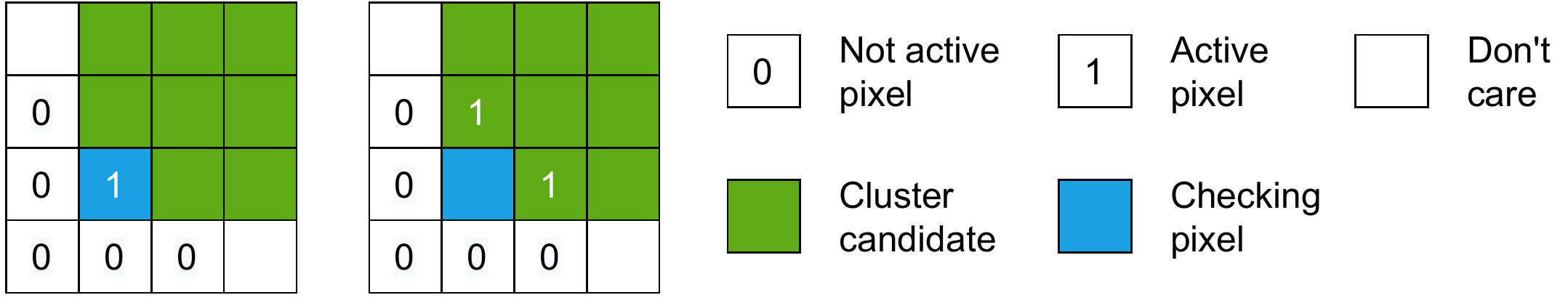}
  \caption{Pixel patterns seeding to a cluster candidate.}
  \label{fig:pattern}
\end{figure}

The pixel geometry of the VELO requires a demanding 2-dimensional cluster-finding task to be performed at the 30 MHz event rate as a preliminary step, before performing track reconstruction.
This has been addressed by a derivation of the same ``Artificial Retina'' approach~\cite{clust}: pixel states are loaded in a matrix of cells, where each cell checks the status of the neighbours looking for specific patterns (Figure~\ref{fig:pattern}). If a cell detects a cluster it returns its position and the status of the pixels in the ``Cluster candidate'' region that will be used to calculate the cluster centre.

Since on average the ratio of active pixels in an event is small ($<$0.01 \%), it is not efficient to implement a cell for each pixel, and we instantiate instead a reduced set of matrices.
VELO data are formatted as 4$\times$2 pixel blocks, named SuperPixels (SPs).
Each matrix can contain up to 9 SPs, in three rows and three columns and it does not map to a specific VELO region until it is initialised.
As a SP arrives to an uninitialised matrix, it fills the matrix in the center, calculating the coordinates of the neighbouring SPs. 
Further SPs input to the matrix are compared with the previously calculated coordinates.
In case of a match, the pixel states are used to fill the right position in the matrix, otherwise the SPs are passed on to the next matrix in the chain.



We implemented the firmware within the already-existing VELO DAQ Boards using roughly 26\% of logics and 10\% of memory of the FPGA.
We tested the firmware on a prototyping board equipped with FPGAs comparable to the ones used at LHCb.
The system runs comfortably without errors at a clock frequency of 350 MHz, providing a measured event rate of 38.9 MHz, amply sufficient to sustain the target rate of 30 MHz readout.

A bit-level simulation of the FPGA clustering algorithm has been implemented and integrated in the official LHCb simulation environment. The HLT tracking is fed with FPGA clusters and its output is compared with that obtained with the standard GPU-based clustering code. The GPU-FPGA comparison has been performed on simulated event samples, at Run-3 conditions (centre of mass energy $\sqrt{s}$ = 14 TeV and luminosity $\mathcal{L} = 2\times10^{33}\,\mathrm{cm}^{-2}~\mathrm{s}^{-1}$).
No significant differences are observed.
This solution is being commissioned as a default component of the LHCb reconstruction already in Run-3.

\section{Distribution network}
\label{sec:network}

VELO forward tracking requires a Tracking Board to be paired to each of the 38 DAQ Boards of the modules downstream of the nominal interaction point, while choosing an appropriate network topology that assures low latency and high bandwidth.
We split the VELO track parameters space in 4 quadrants.
Cells of the same quadrant share a significant number of hits, requiring a highly interconnected network between the corresponding FPGAs.
Cells of different quadrants can also share hits, but they need less bandwidth.

\begin{figure}[htbp]
  \centering
  \begin{subfigure}[b]{0.5\textwidth}
    \centering
    \includegraphics[width=\textwidth]{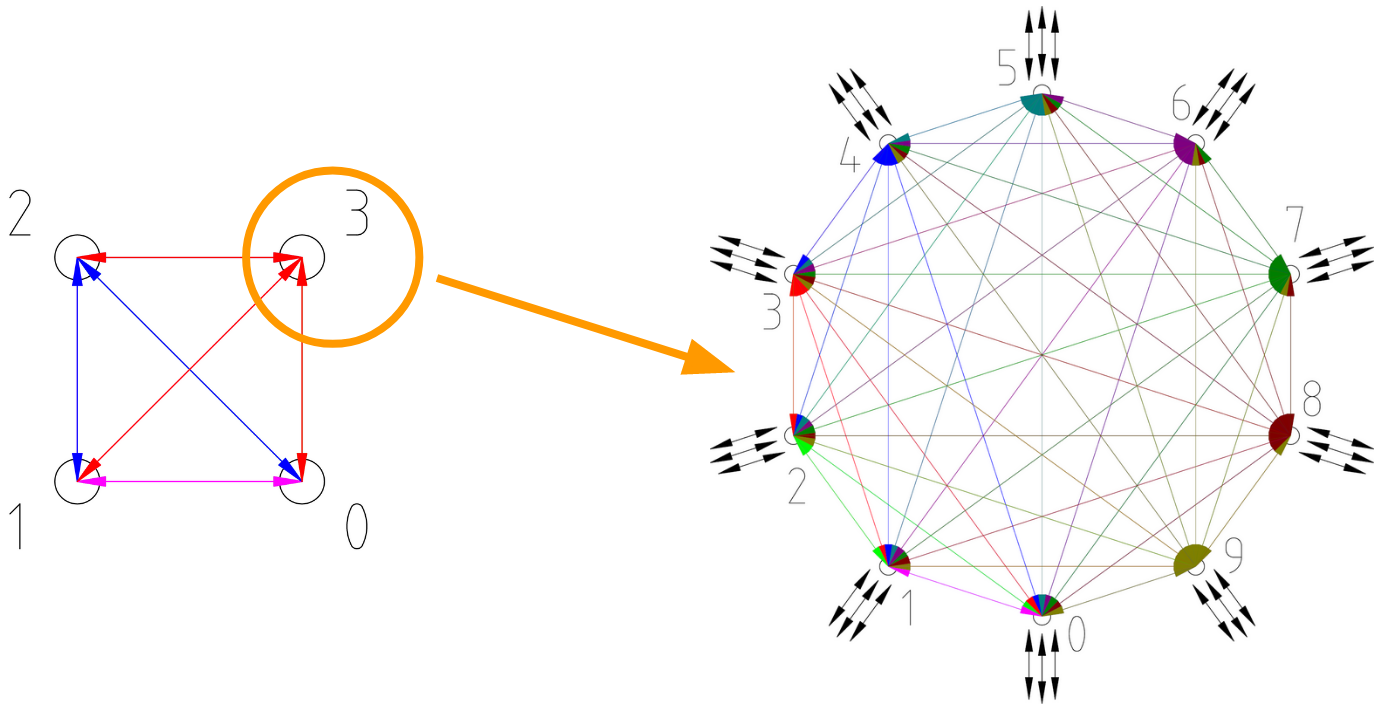}
    \caption{}
    \label{fig:network_topo}
  \end{subfigure}
  \quad
  \begin{subfigure}[b]{0.3\textwidth}
    \centering
    \includegraphics[width=\textwidth]{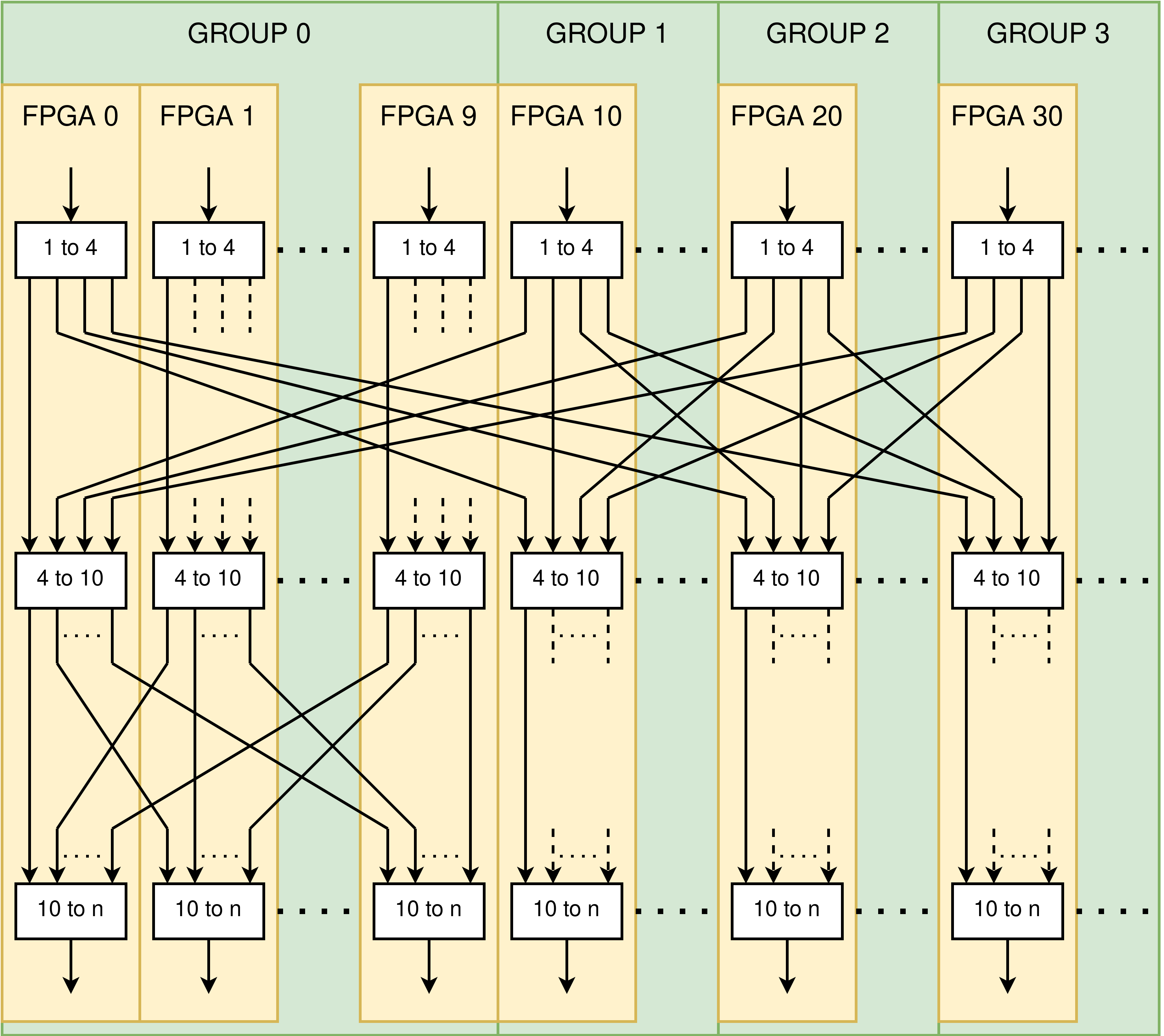}
    \caption{}
    \label{fig:network_detail}
  \end{subfigure}
  \caption{Network topology of a full-mesh network of 4 full-mesh sub-networks based on 10 FPGAs each (a). Details of the Distribution Network developed for the VELO (b).}
  \label{fig:network}
\end{figure}

We can implement 4 full-mesh networks, one for each quadrant, 10 FPGAs each (group).
Then the i-th FPGA of a network is connected to the i-th FPGA of the other three networks creating a full-mesh of full-meshes (Figure~\ref{fig:network_topo}).
Hits exchange between not-directly connected FPGAs require an extra hop, increasing the communication latency, but always remaining below the $\mu$s level.

This configuration requires 12 links for each FPGA.
Several off-the-shelf FPGA boards are equipped with 16 XCVRs, so this topology is implementable without the need for custom hardware.

Figure~\ref{fig:network_detail} shows how we implement the Distribution Network with the described topology.
A set of switches with an input and 4 outputs arrange the data coming from the DAQ Boards by quadrant. The hits relevant to the cell implemented in the same group remain inside the FPGA, while others are sent to the other groups through the inter-group optical links. A second set of switches with 4 inputs and 10 outputs and the intra-group links route the hits to the right FPGA within the group. The final set of switches gather the hits and deliver them to the ``Artificial Retina'' cells implemented in the FPGA.


\subsection{Validation tests}
\label{sec:tests}

We performed tests on the Distribution Network using commercial boards of two different generations.
The older one has an Intel Arria V FPGA and 8 serial links with a maximum bandwidth of 10 Gbps each, the newer one has an Intel Stratix 10 FPGA and 16 high-speed serial links (26 Gbps).
We used the Arria V boards for testing the protocol logic and the connectivity, while  the more advanced Stratix 10 boards are intended for the full-speed prototype in real usage.

At first we validated the communication protocol and the optical hardware generating a pseudo-random sequence in a board and checking it in a different board connected through the optical links. We successfully operated the optical communication on a full-mesh network of 5 Arria V boards, at the maximum speed allowed by the FPGA.

Then we tested a prototype of the Distribution Network spread over the 5 boards. This prototype reproduces all the features of the intra-group stage of the VELO Distribution Network.
Each FPGA contains RAMs pre-loaded with simulated hits, two switches with 4 inputs and 4 outputs, the XCVRs exchanging the switched hits between the boards, and a final component that collects data coming from the other boards, allowing the computer host to verify the correct behaviour of the system.
We tested the system for 5 weeks of continuous operation, with no errors detected.

The same tests are performed on a single Stratix 10 board with the optical fibers closed in loopback.
Also in this case we did not detect errors during the test.
On the Stratix 10 boards we used the SuperLite II V4 communication protocol that allows us to exploit the full potential of the Stratix 10 XCVRs.
Its most useful features for our applications are the low overhead ($>$96 \% of the bandwidth available for payload), and its flow control capability, that we exploit to regulate traffic inside the Distribution Network, avoiding the need for extra buffering.

\section{Conclusion}
\label{sec:conclusion}

We are building a demonstrator that uses the “Artificial Retina” architecture in reconstructing the VELO detector at LHCb. Demonstration of track reconstruction is in advanced state of realisation within the LHCb coprocessor testbed. The first step (cluster finding) is already fully functional, and being commissioned for physics data taking in Run 3. The full Distribution Network topology was defined, and we tested a sizeable prototype of the network.
We hope our efforts will contribute to improving trigger capabilities in future HEP experiments, where a more heterogeneous computing environment will allow achieving an optimal level of performance by assigning the most appropriate device for each task.


\acknowledgments

We are grateful for the funding granted by INFN under CSN1 and CSN5 project RETINA.




\begin{thebibliography}{99}

\bibitem{retina}
R. Cenci et al., \emph{Development of a High-Throughput Tracking Processor on FPGA Boards}, \emph{PoS(TWEPP-17)} {\bf 313} (2018) pg. 136

\bibitem{down}
M. J. Morello et al., \emph{Real-time reconstruction of long-lived particles at LHCb using FPGAs}, \emph{ACAT} (2019)

\bibitem{clust}
G. Bassi et al., \emph{A real-time FPGA-based cluster finding algorithm for LHCb silicon pixel detector}, \emph{EPJ Web Conf.} {\bf 251} (2021) pg. 04016





\end{thebibliography}
\end{document}